\documentclass[journal]{vgtc}                     % final 
\graphicspath{{figures/}{pictures/}{images/}{./}}

\usepackage{times}   
\usepackage{enumerate}

\usepackage{tabu}    
\usepackage{tabularray}
\usepackage{multirow}
\usepackage{graphicx}
\usepackage{adjustbox}
\usepackage{soul}
\usepackage{fancybox}
\usepackage{comment}
\usepackage{amsmath}
\usepackage[table]{xcolor}
\usepackage[normalem]{ulem}

\usepackage{tabu}               
\usepackage{booktabs}           
\usepackage{lipsum}             
\usepackage{mwe}                
\usepackage{mathptmx}           
\usepackage{colortbl}
\usepackage[table]{xcolor}
\usepackage{multirow}
\usepackage{booktabs}
\usepackage{enumitem}
\usepackage[table,xcdraw]{xcolor}
\definecolor{C6EFCE}{HTML}{C6EFCE}
\definecolor{FFEB9C}{HTML}{FFEB9C}
\definecolor{006100}{HTML}{006100}
\definecolor{9C5700}{HTML}{9C5700}

\onlineid{1668}

\vgtccategory{Research}

\vgtcpapertype{application/design study}

\title{Streamlined Facial Data Collection based on Utterance and Emotional Data for Human-to-Avatar Reconstruction}

\author{%
  \fontsize{8.8pt}{10pt}\selectfont
  \authororcid{Seoyoung Kang}{0000-0002-6143-4369}, \authororcid{Seokhwan Yang}{0009-0000-1097-646X}, \authororcid{Hail Song}{0009-0006-4008-196X}, \authororcid{Boram Yoon}{0000-0003-3696-0145}, \authororcid{Jinwook Kim}{0000-0002-1962-5815}, \authororcid{Kangsoo Kim}{0000-0002-0925-378X},  \authororcid{Woontack Woo}{0000-0002-5501-4421}}

\authorfooter{
  \item
  	Seoyoung Kang is with KAIST UVR Lab.
  	E-mail: seoyoung.kang@kaist.ac.kr.
  \item
  	Seokhwan Yang is with KAIST UVR Lab.
  	E-mail: ysshwan147@kaist.ac.kr.
   \item
  	Hail Song is with KAIST UVR Lab.
  	E-mail: hail96@kaist.ac.kr.   
  \item
  	Boram Yoon is with KAIST KI-ITC ARRC.
  	E-mail: boram.yoon1206@kaist.ac.kr.
  \item
  	Jinwook Kim is with KAIST GSCT.
  	E-mail: jinwook.kim31@kaist.ac.kr.
  \item
  	Kangsoo Kim is with the University of Calgary.
  	E-mail: kangsoo.kim@ucalgary.ca.
  \item 
        Woontack Woo is with KAIST UVR Lab and KAIST KI-ITC ARRC. Corresponding Author.
  	E-mail: wwoo@kaist.ac.kr.
}

\abstract{%
  This study explores a streamlined facial data collection method for conversational contexts, addressing the limitations of existing approaches that often require extensive datasets and prioritize technical metrics over user perception and experience. We systematically investigate which facial expression data are essential for reconstructing photorealistic avatars and how they can be captured efficiently. Our research employs a two-phase methodology to identify efficient facial data collection strategies and evaluate their effectiveness. In the first phase, we conduct facial data acquisition and evaluate reconstruction performance using utterance data and emotional data. In the second phase, we carry out a comprehensive user evaluation comparing three progressive conditions: utterance only, utterance and emotional data, and a control condition involving extensive data. Findings from 24 participants engaged in simulated face-to-face conversations reveal that targeted utterance and emotional data achieve comparable levels of perceived realism, naturalness, and telepresence, while reducing training time and data usage when compared to the extensive data collection approach. These results demonstrate that targeted data inputs can enable efficient avatar face reconstruction, offering practical guidelines for real-time applications such as AR/VR telepresence and highlighting the trade-off between data quantity and perceived quality.}


\teaser{
  \centering
 \includegraphics[width=\linewidth]{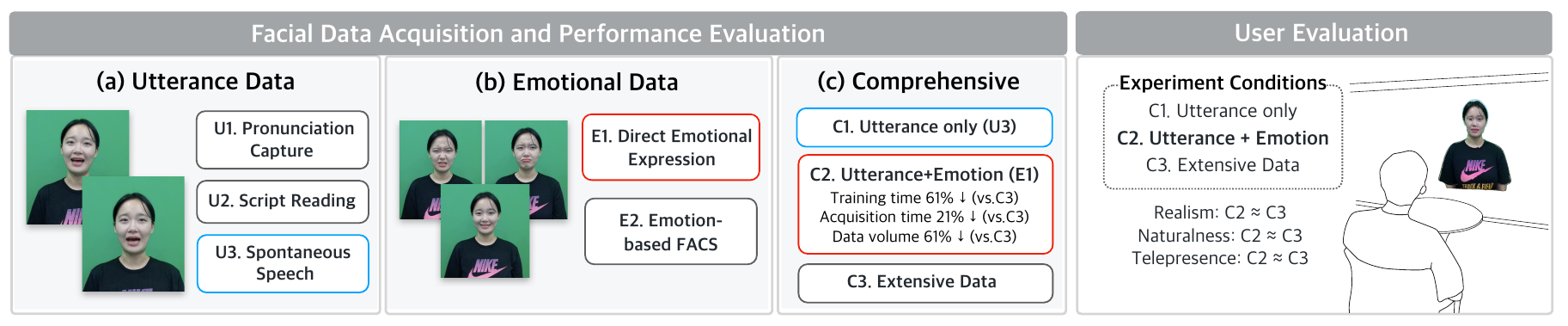}
  \vspace{-3ex}
  \caption{(Left) Optimizing facial data collection by identifying essential expressions: spontaneous speech from (a) utterance data experiment, and direct emotional expression from (b) emotional data experiment were integrated into (c) comprehensive experiment for the Utterance only (C1) and Utterance + Emotion (C2). (Right) The user evaluation simulated a real face-to-face conversation.}
  \label{fig:teaser}
}

\graphicspath{{figs/}{figures/}{pictures/}{images/}{./}} 
\usepackage{tabu}               
\usepackage{booktabs}           
\usepackage{lipsum}              
\usepackage{mwe}                

\usepackage{mathptmx}           

\begin{document}

\firstsection{Introduction}

\maketitle
As Augmented Reality (AR) and Virtual Reality (VR) technologies advance, 3D avatars have become essential for immersive environments, enabling users to express themselves and maintain their identities~\cite{schafer2022survey, kang2025collaboration}.
Demand for realistic and expressive avatars is growing,  particularly in telepresence applications~\cite{orts2016holoportation, stengel2023ai} such as \textit{Project Starline}\footnote{Google Project Starline: \url{https://starline.google/} (Accessed 2026-01-04)}, where authentic interactions are critical. 
However, current methods for generating photorealistic avatars remain computationally intensive, often relying on extensive facial data collection~\cite{lombardi2018deep}. 
Although recent approaches can achieve high technical fidelity, they do not fully address which types of facial data are truly necessary to support effective conversational
interaction.
Therefore, reconstruction processes typically require large datasets and powerful hardware, making them less practical for everyday users~\cite{bai2023learning, chen2023implicit}.

While technical metrics such as accuracy and similarity~\cite{huynh2008scope,wang2004image,zhang2018unreasonable} are essential for evaluating avatars, they do not fully capture the user experience. A key gap in photorealistic avatar research lies in understanding how variations in data collection strategies translate into user perception during communication~\cite{mollahosseini2017affectnet}.
High technical performance does not guarantee user satisfaction; an avatar that looks flawless might still cause discomfort or trigger the uncanny valley~\cite{mori2012uncanny}, ultimately reducing engagement and immersion~\cite{higgins2021ascending}. 
Our study addresses this challenge by examining the minimal yet perceptually adequate facial expression data required for conversational avatar
reconstruction, focusing on two key inputs: utterance and emotional data. Rather than assuming that more data always leads to better results, we emphasize the advantages of a streamlined data collection process. By identifying the optimal combination of utterance and emotional inputs, we aim to demonstrate that high-quality avatars can be built without relying on extensive datasets or excessive computational resources.

Importantly, our research incorporates user evaluation, recognizing that technical metrics alone are insufficient for understanding avatar effectiveness in real-world contexts. By integrating user perceptions of realism, naturalness, communication synchrony, and telepresence, we ensure that avatars are not only efficient to build but also comfortable and engaging for users. Accordingly, our study focuses on the following research questions:

\begin{itemize}[leftmargin=0.4in]
 \setlength\itemsep{-2pt}
    \item [\textbf{RQ1.}] How do different methods of collecting utterance data influence the accuracy and efficiency of avatar facial reconstruction in conversational contexts?

    \item [\textbf{RQ2.}] Which combination of utterance and emotional data results in avatars that maintain high realism while minimizing data collection and computational costs?

    \item [\textbf{RQ3.}] How do different data collection strategies (utterance only, utterance + emotion, and extensive data) shape user perceptions of avatar naturalness, realism, synchrony, and telepresence?
    
     \setlength\itemsep{-2pt}
\end{itemize}

To answer these questions, we designed a two-phase study (Figure \ref{fig:teaser}) aimed at identifying the essential data for high-fidelity avatar reconstruction.
The first phase systematically evaluates facial data acquisition by comparing different utterance (U) and emotional (E) data types to determine the most efficient combination. The second phase serves as a validation study addressing RQ3, comparing three conditions: \textbf{utterance only} (C1), \textbf{utterance + emotion} (C2), and an \textbf{extensive data combination} (C3), to assess how these data collection strategies impact user perceptions in simulated face-to-face conversations. Our findings demonstrate that a strategic combination of spontaneous speech and direct emotion achieves perceptual equivalence to extensive datasets. By reducing data requirements without sacrificing realism, we show that a targeted approach is more effective than simply collecting more data, suggesting that a streamlined data collection strategy can effectively support high-quality avatar communication.

\section{Related Work}

\subsection{Avatars for Telepresence in AR/VR}
Recent advancements in immersive 3D telepresence systems have significantly enhanced the ability of remote AR/VR users to engage in shared environments, supporting seamless communication and collaboration across physical distances~\cite{fuchs2014immersive,orts2016holoportation,10765459}. In these environments, avatars serve as digital representations that allow users to interact in real-time from a first-person perspective~\cite{baker2021avatar}, facilitating telepresence---defined as the psychological sense of being there in an environment by means of a communication medium~\cite{steuer1992defining}. The visual and behavioral realism of these avatars is critical to the quality of communication, as they serve as the primary medium of interaction~\cite{bailenson2004transformed,slater2010first,lee2023effects}. 
Early research introduced generic avatars with customizable features~\cite{combe2024exploring,witherow2024customizable}, but these fell short of reflecting the distinctive and photorealistic features of individual users. 
The shift toward avatars that more accurately reflect real-world appearance has brought new challenges, particularly the time and effort required to capture and process facial expressions. Carrigan et al.~\cite{carrigan2020expression}, for example, proposed expression packing, an efficient blendshape transfer method that minimizes the number of training expressions needed while still supporting personalization.

Recent technological innovations have advanced high-fidelity volumetric rendering techniques, allowing for more efficient capture of dynamic facial expressions~\cite{lombardi2021mixture}. Alongside these developments, Gaussian-splatting methods have emerged to render intricate structures with greater efficiency~\cite{kerbl20233d,moreau2024human,xu2024gaussian}.
On-device computation has also seen significant progress, enabling real-time avatar generation without the need for powerful external hardware. For example, applications like \textit{Apple Vision Pro's Persona (beta)}\footnote{Apple Vision Pro's Persona: \url{https://support.apple.com/guide/apple-vision-pro/dev934d40a17/visionos} (Accessed 2026-01-04)} demonstrate how integrated systems are making avatar creation more accessible for telepresence~\cite{ma2021pixel}. 
Nevertheless, while photorealistic avatars can enhance immersion in telepresence~\cite{fu2023auto}, creating them in real time remains computationally demanding. Capturing and reproducing facial expressions from head-mounted cameras demands substantial computational power, and current pipelines often depend on lengthy and complex data collection sessions. These requirements constrain scalability and make avatar use less practical for everyday users.

Despite the advances, the data collection phase itself often remains a bottleneck. Many approaches still overlook practical concerns such as time and complexity, which can affect adoption among general users. Our work addresses this issue by exploring conversational scenarios and introducing techniques that streamline data capture, reducing both the duration and complexity of the process. By simplifying this stage, our approach lowers barriers to adoption while maintaining realism and expressiveness, ultimately supporting more practical and accessible avatar-based telepresence.

\subsection{Photorealistic Avatars and Face Reconstruction}
Photorealistic avatar face generation has advanced significantly since the introduction of 3D Morphable Models (3DMM)~\cite{blanz2023morphable}, which used principal component analysis (PCA) to represent facial appearance and expressions in a low-dimensional subspace. 
By simplifying the complexity of facial models into manageable vectors, 3DMM became the cornerstone for facial reconstruction techniques and has been expanded with parametric head models like FLAME~\cite{li2017learning}, incorporating expressions, shape, and pose for more comprehensive face representations. Neural Radiance Fields (NeRF) have revolutionized avatar reconstruction, enabling high-quality 3D Scenes from limited inputs like monocular RGB videos~\cite{gafni2021dynamic, grassal2022neural, zheng2022avatar}. 
Gafni et al.~\cite{gafni2021dynamic} introduced Dynamic NeRF for 4D avatars, improving movement and viewpoint consistency. Further improvements by Grassal et al.~\cite{grassal2022neural} and Zheng et al.~\cite{zheng2022avatar} refined neural head avatars with implicit morphable models, increasing realism and versatility. Recent works such as Gaussian Head Avatar~\cite{xu2024gaussian} and HyperGaussians~\cite{Serifi2025HyperGaussians} have advanced avatar representations with greater realism, while methods like RGBAvatar~\cite{li2025rgbavatar}, SplattingAvatar~\cite{shao2024splattingavatar}, and FlashAvatar~\cite{xiang2024flashavatar} have focused on reducing computational demands while maintaining high-fidelity avatars.

Despite these technical advances, a key gap remains in understanding how these improvements impact user perception. Most research focuses on technical metrics such as accuracy and similarity ~\cite{huynh2008scope,wang2004image,zhang2018unreasonable}, often overlooking how these translate into real-world applications in terms of perceived realism, naturalness, and expressiveness for avatar-based interactions. Subtle gains in fidelity may not meaningfully enhance the overall user experience, leaving a disconnect between high technical performance and user expectations. 
Our approach addresses this by focusing on everyday, conversation-driven scenarios where the efficiency of avatar creation is more crucial. By prioritizing essential inputs such as utterance and emotional data, we generate expressive, realistic avatars while reducing computational demands. This balance of technical performance and user experience makes our avatars ideal for scalable, real-time AR/VR applications, ensuring immersive communication without excessive resource use.

\section{Methodology}

This study examines the optimization of data collection for photorealistic avatar face reconstruction by identifying essential inputs that balance realism, expressiveness, and computational efficiency. We adopted a two-phase design: Phase 1 evaluated the performance of different methods for capturing utterance and emotional data, while Phase 2 compared user perceptions of avatars generated from these methods. This structure enables us to assess whether reductions in data complexity lead to perceptible differences in avatar quality, thereby identifying cost-efficient strategies without compromising user experience.

\subsection{Facial Data Acquisition and Evaluation}
\label{collection}
Our study focuses on investigating two key inputs, \textbf{utterance data} and \textbf{emotional data}, to determine which facial expressions are essential for conversational avatars. We aimed to test different acquisition methods and identify the most effective techniques for capturing natural and expressive facial movements.

\subsubsection{Utterance Data Acquisition}
Utterance-related expressions are fundamental in generating avatars that can accurately replicate the facial dynamics of everyday conversational behaviors~\cite{pelachaud1996generating}. Utterance data refers to speech-driven facial movements captured during active speech, reflecting coordinated articulatory dynamics associated with phoneme production~\cite{harris1963structural}. 
All three methods included baseline recordings of head and eye movements to capture essential conversational dynamics. To explore different acquisition strategies, we tested three methods:

\begin{itemize}
\setlength\itemsep{0pt}
    \item {\textbf{Pronunciation Capture (U1):}} 
    Participants articulated key Korean consonants and vowels (e.g., ``Ga-Na-Da-Ra" and ``A-E-I-O-U"). This method examined whether a minimal dataset of core articulations could yield realistic conversational avatars.

    \item {\textbf{Script Reading (U2):}} Participants read aloud pre-written scripts consisting of news articles, capturing structured and consistent speech patterns. A script was provided to ensure uniformity across participants. This controlled approach helped us assess whether standardized data improves reconstruction quality.

    \item {\textbf{Spontaneous Speech (U3):}} Participants delivered unscripted, informative speeches on freely chosen topics, producing more dynamic and naturally varying expressions. This approach captures the natural flow of conversation, investigating how well avatars trained on free-topic, spontaneous speech data captured real-world variability.
\end{itemize}

\subsubsection{Emotional Data Acquisition}
To complement the utterance data, we conducted emotional expression experiments. Each utterance method was paired with two emotion-capture approaches, Direct Emotional Expression and Emotion-based FACS, resulting in six data-collection strategies: U1E1, U1E2, U2E1, U2E2, U3E1, and U3E2.

\begin{itemize}
\setlength\itemsep{0pt}
    \item {\textbf{Direct Emotional Expression (E1):}} Participants expressed basic emotions~\cite{ekman1970universal}, using Paul Ekman’s facial expression images as visual references, which guided them in producing the target expressions. The emotions included happiness, sadness, anger, fear, disgust, surprise, and neutral, capturing immediate and direct facial expressions~\cite{ekman1976pictures}.

    \item {\textbf{Emotion-based FACS (E2):}} 
    Participants reproduced Facial Action Units (AUs) associated with basic emotions~\cite{ekman1978facial,friesen1983emfacs}, following GIF animations as guides. This method targeted the fine-grained muscle movements across the brows, eyes, nose, cheeks, lips, and jaw, capturing the subtle dynamics that underlie the emotional expressions.

    \setlength\itemsep{-2pt}
\end{itemize}

\subsection{Comprehensive Experiment}
\label{comprehensive}
The second phase of our research compared three conditions based on the most effective acquisition methods from Phase 1 (described in Section \ref{collection}). This experiment allowed us to test whether streamlined approaches could perform comparably to data-intensive methods in both technical and user-centered evaluations.

\begin{itemize}
\setlength\itemsep{0pt}
    \item {\textbf{Utterance only (C1):}} 
    Based on the utterance data experiment results, we used the most effective for capturing natural facial expressions. This condition evaluated whether avatars generated using only utterance data could achieve sufficient realism and naturalness in conversational interactions.
    \item {\textbf{Utterance and Emotion (C2):}} 
    This condition combined the best-performing methods from the utterance and emotional data experiment. We tested whether this optimal combination enhanced avatar performance while maintaining computational efficiency.
    \item {\textbf{Extensive Data Combination (C3):}}
    This condition incorporated a comprehensive range of facial expressions beyond the focused methods used in C1 and C2. It served as a baseline for comparison, helping us evaluate whether collecting substantially more data provided meaningful improvements over the streamlined approaches.
    
\end{itemize}

\begin{figure}[t]
    \centering
    \includegraphics[width=\linewidth]{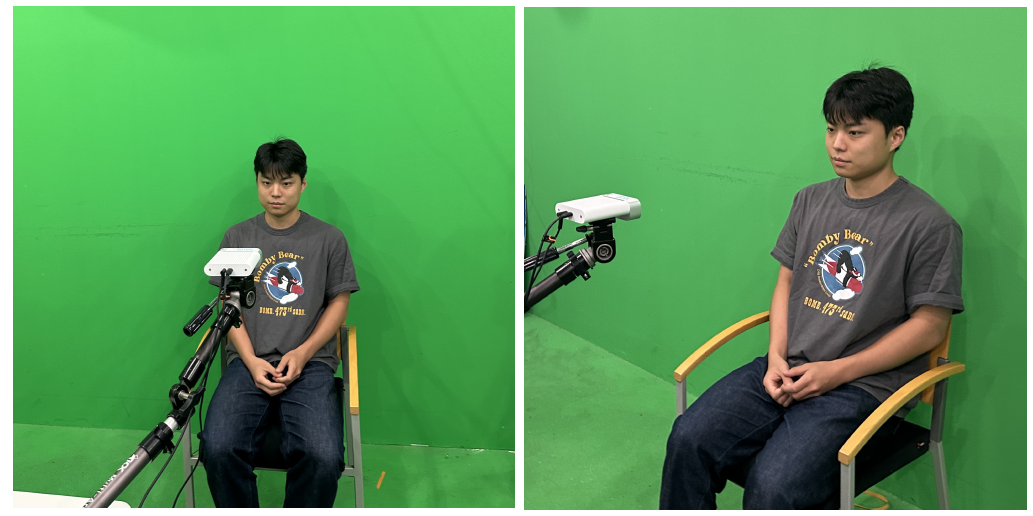}
    \caption{(Left) Data collection setup showing the front-facing participant from the front view and (Right) the side view.}
    \label{Fig:Results}
\end{figure}

\subsection{Apparatus}
Data were collected with the Microsoft Azure Kinect\footnote{Microsoft Azure Kinect: \url{https://learn.microsoft.com/en-us/azure/kinect-dk/} (Accessed 2026-01-04)}, positioned 0.6 meters in front of participants to capture clear facial movements (as shown in Figure~\ref{Fig:Results}).
While the device supports both RGB and depth, only RGB video was recorded to match the requirements of our reconstruction pipeline~\cite{zheng2023pointavatar}. All videos were originally captured at a resolution of 3840 × 2160 pixels with a frame rate of 30 fps, and were later resized to 512 × 512 pixels during preprocessing to meet the input requirements of the reconstruction pipeline. The camera was calibrated to ensure consistent and accurate data capture across all sessions. 
We recorded facial expression video data and combined them to generate each experimental condition, which was then used to reconstruct the avatar’s face from the video frames. Additionally, we recorded test videos for both quantitative evaluation and the user study. These videos captured conversational scenarios simulating everyday communication, enabling us to assess the avatar’s reconstruction performance. 

Video acquisition was performed on a desktop system equipped with an NVIDIA GeForce RTX 3090 GPU (24 GB VRAM) and an Intel(R) Core(TM) i9-10900K CPU, which was used exclusively for data recording and storage. The recorded videos were subsequently transferred to a server for avatar reconstruction and further processing. Avatar reconstruction and rendering were conducted on a server equipped with an NVIDIA A40 GPU (40 GB VRAM) and an Intel Xeon Gold 6326 CPU. During the reconstruction process, up to approximately 98\% of the available GPU VRAM was allocated, whereas system memory usage remained low, with at most about 5\% of the total 256 GB RAM utilized.

\subsection{Procedure}
After collecting the datasets under diverse conditions, we reconstructed avatars using PointAvatar~\cite{zheng2023pointavatar}, a neural point cloud–based method well suited for high-fidelity expression modeling in conversational settings. In accordance with the PointAvatar framework, which operates solely on RGB video inputs, depth data was not recorded.
The reconstruction process began with preprocessing, which involved background removal, segmentation, and video normalization. All videos were resized to a resolution of 512 × 512 pixels, and frame lengths were standardized to match the longest sequence in the dataset. These steps ensured that the data were both clean and comparable across conditions. To construct experimental conditions, we leveraged PointAvatar’s configuration system to systematically include or exclude specific data components during training.
By selectively incorporating utterance or emotion data, we ensured that only the data composition changed, while temporal synchronization and computational environments remained identical, enabling a controlled ablation study that preserves natural conversational dynamics.

Next, parametric modeling was performed using FLAME~\cite{li2017learning} to extract facial structures and expressions from the preprocessed frames. These parameters were then applied to deform canonical point clouds, allowing us to replicate a full range of facial movements. The deformed point clouds were then rendered into high-quality images. This stage applied shading and detail preservation techniques to capture the fine-grained facial details, resulting in high-quality rendered outputs. Finally, the reconstructed frames were integrated into the Unity environment. Each frame sequence was exported as a .ply file and configured to play back at 30 frames per second, ensuring temporal consistency. This step enabled smooth animation and realistic avatar representation for evaluation in conversational scenarios. Together, this pipeline provided a consistent and rigorous framework for comparing different data collection strategies across both quantitative analyses and user evaluations.

\subsection{Measures}

\subsubsection{Performance Evaluation}
To evaluate the performance of different methods, we assessed both reconstruction quality and computational cost using quantitative image comparison metrics and computational cost measures. This dual approach enabled a comprehensive comparison of reconstruction quality as well as the efficiency and feasibility of each method.

For image comparison, we employed four widely used metrics: MAE (Mean Absolute Error), SSIM (Structural Similarity Index Measure)~\cite{wang2004image}, PSNR (Peak Signal-to-Noise Ratio)~\cite{huynh2008scope}, and LPIPS (Learned Perceptual Image Patch Similarity)~\cite{zhang2018unreasonable}. MAE captures pixel-level accuracy by measuring the average absolute difference between the rendered and original images. SSIM evaluates how well the structural details of the original image are preserved, with higher scores indicating greater similarity. PSNR measures image quality in terms of signal-to-noise ratio, where higher values indicate clearer, less noisy reconstructions. LPIPS assesses perceptual similarity based on deep learning features, with lower scores representing closer perceptual alignment with the original.

For computational costs, we measured Training Time, Data Acquisition Time, and the Amount of Data collected. Training Time recorded the duration required to train each avatar model. Data Acquisition Time captured the time spent collecting facial video data for each condition. The Amount of Data was measured by the total file size of the captured videos, providing a practical view of storage and processing requirements.

\subsubsection{User Evaluation}
\label{userstudymetric}
In addition to performance metrics, we conducted a user study to capture both quantitative ratings and qualitative feedback on avatars generated with different data collection methods. The primary objective was to understand how users perceived these avatars and to identify factors that influence the overall experience. 
Adapted from Fraser et al.~\cite{fraser2022expressiveness}, we evaluated Facial Animation Realism to assess the avatar’s facial expressions and movements. We utilized this scale to measure realism (“The facial expressions of the avatar looked realistic”) and naturalness (“The facial expressions of the avatar looked natural”).
To facilitate a more detailed analysis of user perception, we analyzed realism and naturalness as distinct sub-dimensions.
We also assessed telepresence~\cite{nowak2003effect} with questions such as, “How involving was the experience?” and “To what extent did you feel immersed in the environment you saw/heard?” Finally, drawing upon Kang et al.~\cite{kang2024influence}, we evaluated synchrony of communication cues, examining how well the avatar’s expressions aligned with the intended conversational signals.
Since the avatar interactions (stimuli) were one-way and identical across conditions, we excluded questions related to communication efficacy, which could have been affected by repetitive speeches. Instead, the study focused on naturalness, realism, synchrony of communication cues, and telepresence, investigating whether avatars felt engaging, believable, and immersive in conversational scenarios.

\section{Study I: Facial Data Acquisition and Performance Evaluation}
\label{section}

Following the data collection phases described in Section \ref{collection}, we reconstructed 3D avatars using PointAvatar~\cite{zheng2023pointavatar} to replicate participants’ facial features and expressions. In each experiment, all sequences were resampled to match the length of the longest sequence to ensure comparable computational load across conditions. The goal of Study 1 was to identify the most effective data acquisition strategies for generating realistic avatars. Accordingly, we conducted two experiments:

\begin{itemize}
    \item \textbf{Utterance Data Experiment:} testing three methods of capturing facial expressions during speech: pronunciation capture (U1), script reading (U2), and spontaneous speech (U3).
    
    \item \textbf{Utterance and Emotional Data Experiment:} combining each utterance method with two emotional data acquisition techniques: direct emotional expression (E1) and emotion-based FACS (E2). This resulted in six conditions (U1E1, U1E2, U2E1, U2E2, U3E1, U3E2) for evaluation.
\end{itemize}

\begin{table}[t]
\caption{Performance evaluation in the Utterance data experiment, comparing three data collection methods: Pronunciation Capture (U1), Script Reading (U2), and Spontaneous Speech (U3).}
\label{table1}
\centering
\renewcommand{\arraystretch}{1.2}
\resizebox{\columnwidth}{!}{
\begin{tabular}{cccccc}
\toprule
\textbf{Subject} & \textbf{Method} & \textbf{MAE $\downarrow$} & \textbf{PSNR $\uparrow$} & \textbf{SSIM $\uparrow$} & \textbf{LPIPS $\downarrow$} \\ \midrule
\multirow{3}{*}{male\_1} 
 & U1 & 0.0048 & 30.2 & 0.969 & 0.0434 \\
 & U2 & 0.0046 & 30.6 & 0.966 & 0.0511 \\
 & U3 & \cellcolor{green!25}0.0046 & \cellcolor{green!25}31.3 & \cellcolor{green!25}0.969 & \cellcolor{green!25}0.0425 \\ \hline
 
\multirow{3}{*}{female\_1} 
 & U1 & 0.0104 & 24.4 & 0.958 & 0.0562 \\
 & U2 & 0.0092 & 25.2 & 0.963 & 0.0480 \\
 & U3 & \cellcolor{green!25}0.0083 & \cellcolor{green!25}26.0 & \cellcolor{green!25}0.966 & \cellcolor{green!25}0.0447 \\ \hline
 
\multirow{3}{*}{male\_2} 
 & U1 & 0.0040 & 33.6 & 0.985 & 0.0111 \\
 & U2 & 0.0026 & 36.4 & 0.989 & 0.0093 \\
 & U3 & \cellcolor{green!25}0.0023 & \cellcolor{green!25}37.7 & \cellcolor{green!25}0.990 & \cellcolor{green!25}0.0086 \\ \hline
 
\multirow{3}{*}{female\_2} 
 & U1 & 0.0070 & 28.2 & 0.972 & 0.0291 \\
 & U2 & 0.0059 & 29.5 & 0.975 & 0.0247 \\
 & U3 & \cellcolor{green!25}0.0051 & \cellcolor{green!25}30.3 & \cellcolor{green!25}0.980 & \cellcolor{green!25}0.0219 \\ \hline

 \end{tabular}}
\end{table}

\subsection{Phase 1: Utterance Data Experiment}
\label{utt}

Table~\ref{table1} reports the reconstruction performance of U1–U3 across four subjects \((\text{average} \pm \text{std})\) using MAE, PSNR, SSIM, and LPIPS. Pairwise t-tests were used to examine statistical differences between methods.
\if 0 
presents the quantitative evaluation of avatar face reconstruction across three data collection methods: Pronunciation Capture (U1), Script Reading (U2), and Spontaneous Speech (U3). Below, the performance metrics—MAE, PSNR, SSIM, and LPIPS—are reported alongside their mean and standard deviation across four subjects (reported as \((\text{average} \pm \text{std})\). The results are complemented by pairwise t-tests to assess statistical significance.
\fi 
Overall, U3 (Spontaneous Speech) consistently achieved the best performance across all metrics. It showed the lowest reconstruction error (MAE = \((0.0051 \pm 0.0025)\)), reflecting higher accuracy and more stable results across participants. 
Pairwise analysis confirmed that U3 significantly outperformed U1 (\(t = 3.72, p < 0.05\)), while the difference between U3 and U2 was not statistically significant (\(t = 2.24, p = 0.113\)).
In terms of image quality, U3 achieved the highest PSNR (PSNR = \((31.33 \pm 4.82)\)), with significant improvements over U1 (\(t = -3.58, p < 0.05\)) but only marginal differences compared to U2 (\(t = -2.48, p = 0.089\)).
U3 also obtained the highest structural similarity (SSIM = \((0.976 \pm 0.010)\)), indicating better structural alignment, and significantly outperformed U1 (\(t = 4.15, p < 0.05\)). Finally, U3 yielded the lowest perceptual error (LPIPS = \((0.0294 \pm 0.0161)\)), showing the closest perceptual match to original images.
Pairwise tests showed that U3 significantly outperformed U1 (\(t = -3.85, p < 0.05\)), with marginal but statistically non-significant advantages over U2 (\(t = -2.52, p = 0.088\)).

%but shows smaller, statistically non-significant improvements over U2 (\(t = 1.92, p = 0.141\)).
To sum up, U3 emerged as the most effective utterance method, achieving significantly higher accuracy and perceptual quality than U1. Compared to U2, U3 also demonstrated consistently stronger performance across all measures, although the differences did not always reach statistical significance.

\subsection{Phase 2: Utterance and Emotional Data Experiment}
\label{emo}

Table~\ref{tab:table3} shows the results for six combinations of utterance and emotional data collection methods.
The U3E1 condition (Spontaneous Speech + Direct Emotional Expression) achieved the strongest overall performance. It produced the lowest reconstruction error (MAE = \((0.0049 \pm 0.0024)\)) and significantly outperformed both U1E1 (\(t = 3.85, p < 0.05\)) and U1E2 (\(t = 4.02, p < 0.05\)). Although U3E1 also outperformed U2E1 and U2E2, these differences were not statistically significant. U3E1 further achieved the highest PSNR \((31.65 \pm 4.96)\)), significantly surpassing U1-based conditions (U1E1 (\(t = -3.82, p < 0.05\)) and U1E2 (\(t = -3.95, p < 0.05\))), and reached the highest SSIM (SSIM = \((0.978 \pm 0.0101)\)), again with significant improvements over U1 combinations. Its perceptual similarity was also strongest (LPIPS = \((0.0282 \pm 0.016)\), with clear advantages over U1 conditions and marginal improvements over U2.
Overall, these results indicate that combining spontaneous speech with direct emotional expression is the most effective strategy for capturing rich and natural facial dynamics.
Across all objective metrics, U3E1 significantly outperformed U1-based methods and showed clear trends toward improved performance compared to U2. The results further suggest that direct emotional expression (E1) generally yields better outcomes than FACS-based expression (E2), particularly when paired with spontaneous speech.

\subsection{Comprehensive Experiment}
We additionally evaluated the computational costs associated with each facial data collection condition: C1 (utterance only), C2 (utterance + emotion), and C3 (extensive data combination), using data averaged across four participants. Three key efficiency-related metrics were considered: training time, acquisition time, and amount of data collected.
As shown in Table \ref{tab:table6}, the results show differences in computational and data efficiency across the three conditions. 

C1 required the least resources, with a training time of 34 minutes 56 seconds, an acquisition time of 60.8 seconds, and a data volume of 209.36 MB. C2 took 44 minutes 24 seconds to train, 77.0 seconds for acquisition, and generated 264.11 MB of data. C3, which involved extensive data collection, was the most resource-intensive, with a training time of 1 hour 53 minutes 2 seconds, an acquisition time of 197.5 seconds, and 682.25 MB of data.

\begin{table}[t]
\caption{
Avatar facial reconstruction performance across six data collection methods. Methods combine utterance data (U1: Pronunciation Capture, U2: Script Reading, U3: Spontaneous Speech) with emotional expression data (E1: Direct Expression, E2: FACS-based Emotion), resulting in six combinations (U1E1, U1E2, U2E1, U2E2, U3E1, U3E2)}
\vspace{-1ex}
\label{tab:table3}
\centering
\renewcommand{\arraystretch}{1.2}
\resizebox{\columnwidth}{!}{
\begin{tabular}{cccccc}
\toprule
\textbf{Subject} & \textbf{Method} & \textbf{MAE $\downarrow$} & \textbf{PSNR $\uparrow$} & \textbf{SSIM $\uparrow$} & \textbf{LPIPS $\downarrow$} \\ \midrule
\multirow{6}{*}{male\_1} 
 & U1E1 & 0.0051 & 30.5 & 0.969 & 0.0410 \\ 
 & U1E2 & 0.0049 & 30.4 & 0.967 & 0.0466 \\ 
 & U2E1 & 0.0051 & 30.5 & 0.967 & 0.0434 \\ 
 & U2E2 & 0.0052 & 30.0 & 0.964 & 0.0496 \\ 
 & U3E1 & \cellcolor{green!25}0.0044 & \cellcolor{green!25}31.7 & \cellcolor{green!25}0.972 & \cellcolor{green!25}0.0397 \\ 
 & U3E2 & 0.0047 & 31.2 & 0.969 & 0.0443 \\ 
 \hline
 
\multirow{6}{*}{female\_1} 
 & U1E1 & 0.0093 & 25.1 & 0.963 & 0.0501 \\ 
 & U1E2 & 0.0099 & 24.7 & 0.960 & 0.0536 \\ 
 & U2E1 & 0.0086 & 25.7 & 0.967 & 0.0459 \\ 
 & U2E2 & 0.0087 & 25.6 & 0.965 & 0.0470 \\ 
 & U3E1 & \cellcolor{green!25}0.0080 & \cellcolor{green!25}26.3 & \cellcolor{green!25}0.968 & \cellcolor{green!25}0.0431 \\ 
 & U3E2 & 0.0085 & 25.8 & 0.966 & 0.0443 \\ 
 \hline
 
\multirow{6}{*}{male\_2} 
 & U1E1 & 0.0027 & 35.5 & 0.988 & 0.0103 \\ 
 & U1E2 & 0.0031 & 35.1 & 0.987 & 0.0102 \\ 
 & U2E1 & 0.0028 & 35.9 & 0.988 & 0.0090 \\ 
 & U2E2 & 0.0026 & 36.7 & 0.989 & 0.0087 \\ 
 & U3E1 & \cellcolor{green!25}0.0021 & \cellcolor{green!25}38.2 & \cellcolor{green!25}0.991 & \cellcolor{green!25}0.0079 \\ 
 & U3E2 & 0.0025 & 37.1 & 0.990 & 0.0087 \\
 \hline
 
\multirow{6}{*}{female\_2} 
 & U1E1 & 0.0066 & 28.5 & 0.973 & 0.0280 \\ 
 & U1E2 & 0.0068 & 28.3 & 0.973 & 0.0287 \\ 
 & U2E1 & 0.0060 & 29.4 & 0.975 & 0.0253 \\ 
 & U2E2 & 0.0061 & 29.2 & 0.974 & 0.0263 \\ 
 & U3E1 & \cellcolor{green!25}0.0050 & \cellcolor{green!25}30.4 & \cellcolor{green!25}0.980 & \cellcolor{green!25}0.0221 \\ 
 & U3E2 & 0.0054 & 30.1 & 0.979 & 0.0225 \\ 
 \bottomrule

 \end{tabular}}
 % \vspace{-1ex}
\end{table}

\begin{table}[t!]
\caption{Comparison of computational costs across conditions, averaging data from four participants. Training time refers to the total time required to train the model. Acquisition time indicates the duration needed to gather data, and the amount of data represents the volume collected for each condition.}
\vspace{-1ex}
\label{tab:table6}
\centering
\renewcommand{\arraystretch}{1.3}
\resizebox{\columnwidth}{!}{
\begin{tabular}{cccc}
\toprule
\textbf{Method} & \textbf{Training Time} & \textbf{Acquisition Time} & \textbf{Amount of Data} \\ 
\midrule
C1 & \cellcolor{green!25}34m 56s &  \cellcolor{green!25}60.8 seconds &  \cellcolor{green!25}209.36 MB \\
C2 & \cellcolor{green!25}44m 24s & \cellcolor{green!25}77.0 seconds & \cellcolor{green!25}264.11 MB \\
C3 & 1h 53m 2s & 197.5 seconds & 682.25 MB \\ \bottomrule
\end{tabular}
% \vspace{-1ex}
}
\end{table}

\begin{figure}[t!]
\centering
\includegraphics[width=0.8\linewidth]{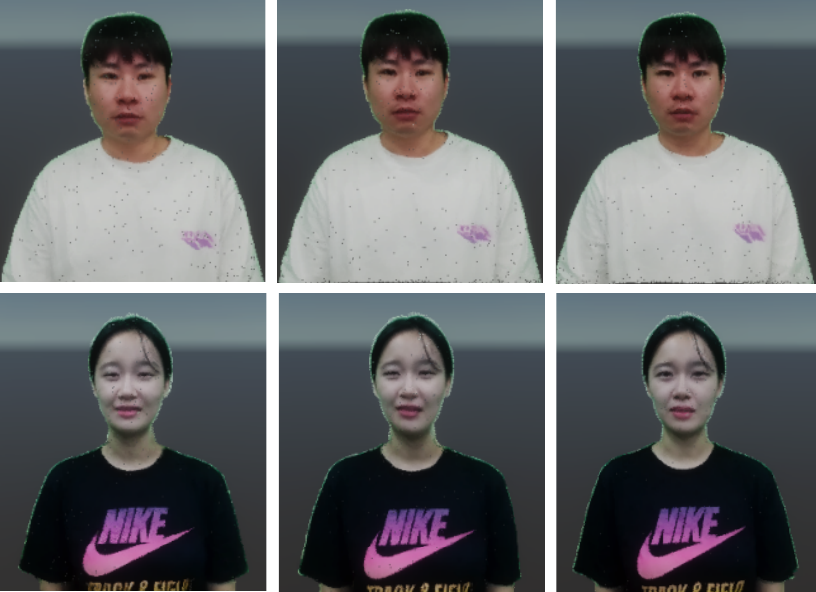}
% \vspace{-1ex}
\caption{Unity view of the avatars used in the user evaluation. (Top) Male Avatar, (Bottom) Female Avatar. (Left) Utterance only, (Middle) Utterance and Emotion, (Right) Extensive Data Combination.}
\label{Fig:unity}
% \vspace{-1ex}
\end{figure}

\section{Study II: User Evaluation}
\label{Sec:Comprehensive}
Addressing RQ3, we investigated how different facial data collection approaches affect people’s perceptions of avatars during conversational interactions, focusing on four key aspects of user experience: realism, naturalness, synchrony of communication cues, and telepresence. 
While Study 1 (described in Section \ref{section}) mainly assessed technical performance using metrics such as MAE, PSNR, SSIM, and LPIPS, these measures alone cannot capture how users perceive avatars in conversation. To address this, we conducted a user study that emphasized the human perspective, collecting both quantitative ratings and qualitative feedback to evaluate how participants perceived avatars generated from three different data collection methods.

\if 0 
Our user study investigated how different approaches to collecting facial data affect people's perceptions when interacting with avatar faces in conversations. We specifically examined four key aspects of user experience:  realism, naturalness, synchrony of communication cues, and telepresence. 

While we had evaluated the technical quality of our avatars using standard metrics (MAE, PSNR, SSIM, and LPIPS) in Study 1, we recognized that these measurements alone could not tell us how users would actually experience and interact with the avatars in real conversations. Therefore, we conducted a comprehensive user study to understand how people perceive and respond to different versions of our avatars during face-to-face interactions, focusing on the human experience rather than just technical performance.
\fi

\subsection{Study Design}
\label{studydesign}
We implemented a within-subject design in which each participant experienced all three avatar conditions: 

\begin{itemize}
    \item \textbf{C1:} utterance only,
    \item \textbf{C2:} utterance + emotion,
    \item \textbf{C3:} extensive data combination.
\end{itemize}

This approach allowed direct comparison of the avatars while controlling for individual differences. Participants provided both quantitative ratings and qualitative feedback, offering insights into the strengths and weaknesses of each condition and guiding directions for future avatar development.

\subsection{Experimental Setup}
The study was conducted in an immersive environment featuring a large-scale projection on a wall display powered by Samsung’s Premiere 9 short-throw projector (4K: 3840$\times$2160, max width: $330\,cm$, throw ratio: 0.189). 
When fully utilized, the display projected an area of approximately 420 cm by 236 cm, covering 78.6\% of the full 4K resolution.
To enhance immersion, the experimental space (4.2 meters by 4 meters) was enclosed with blackout curtains and fully darkened. Avatars were displayed at a life-sized scale and positioned 2 meters from participants, corresponding to the typical social distance~\cite{hall1968proxemics}.

We chose the wall display over head-mounted devices to ensure a stable operational environment for the high-fidelity avatars, maintaining consistent resolution and visual quality while eliminating hardware-induced variability. While our avatars are reconstructed as deformable 3D point clouds using PointAvatar~\cite{zheng2023pointavatar}, we presented them as fixed-viewpoint sequences (Figure~\ref{Fig:unity}) to ensure visual stability. Regarding auditory stimuli, pre-recorded speech was delivered through speakers synchronized with the facial animations. The system ran on a desktop with an NVIDIA GeForce RTX 3090 GPU and an Intel Core i9-10900K CPU.

\begin{figure*}[t]
%\vspace{-1ex}
 \centering
 \includegraphics[width=0.9\linewidth]{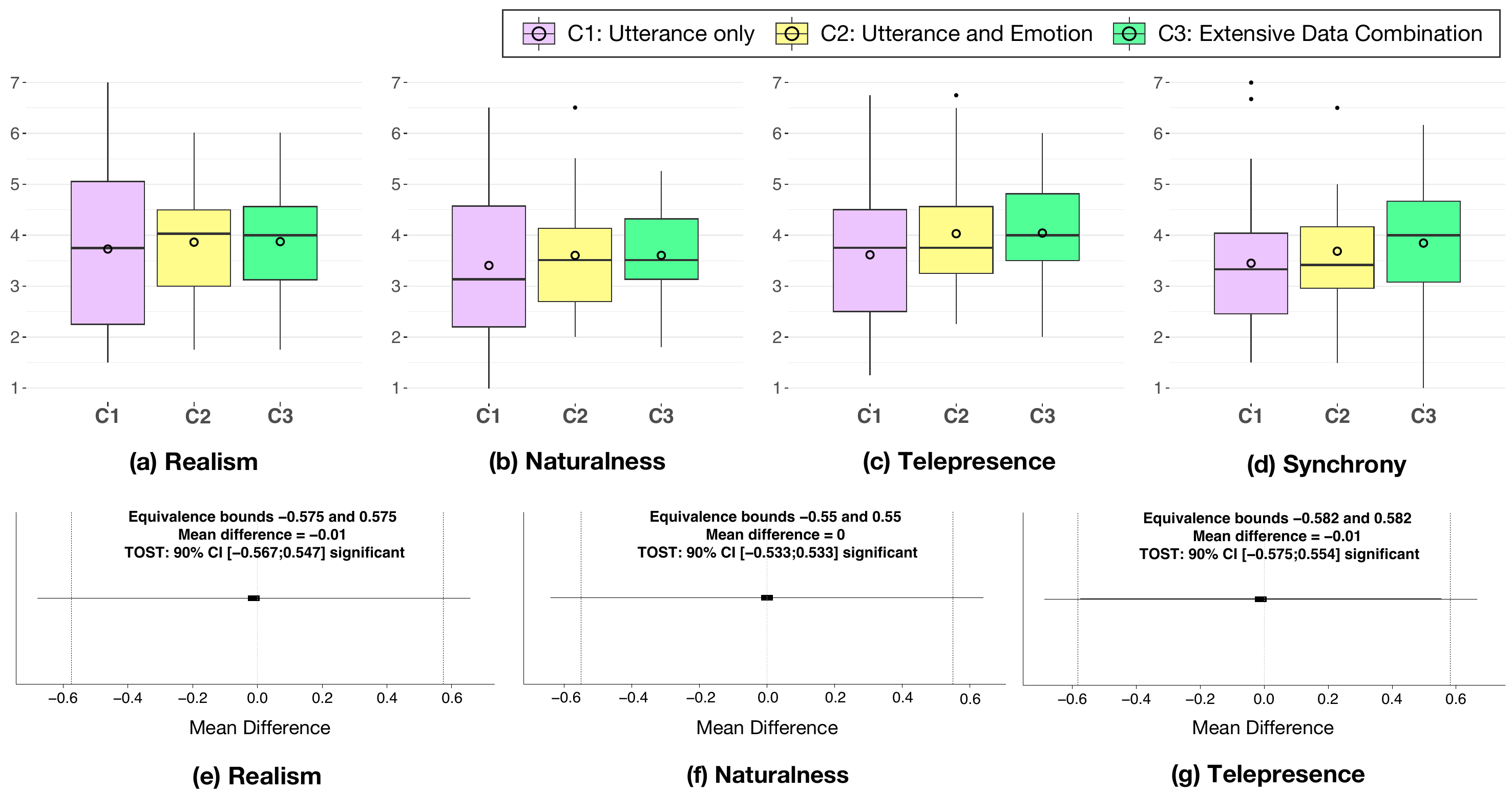}
 %\vspace{-1ex}
 \caption{(Top) Results of Likert scale ratings (1: strongly disagree – 7: strongly agree) for (a) Realism, (b) Naturalness, (c) Telepresence, and (d) Synchrony of Communication cues. (Bottom) Equivalence testing results comparing C2 and C3 for (e) Realism, (f) Naturalness, and (g) Telepresence.} 
 \label{fig:results}
\end{figure*}

\subsection{Participants}
For the main experiment, a new group of 24 participants (12 males and 12 females, $M = 27.8$, $SD = 4.36$) was recruited to evaluate the reconstructed avatars. All participants had normal or corrected-to-normal vision and hearing, ensuring their full engagement with the immersive environment and avatar interactions. All participants self-identified as Asian, reflecting the demographic background of the subject used for avatar reconstruction and ensuring consistency between the reconstructed avatar and the evaluators. All participants reported varying levels of prior VR/AR experience, with 13 participants (54.2\%) having engaged in 10 or more VR/AR sessions, 5 participants (20.8\%) reporting 5–10 sessions, 1 participant (4.2\%) reporting 3–5 sessions, 3 participants (12.5\%) reporting 1–3 sessions, and 2 participants (8.3\%) reporting no prior experience. Regarding specific experiences, 22 participants (91.7\%) had prior experience with 2D video-based remote collaboration, 21 participants (87.5\%) had experience with avatar-based video games, and 5 participants (20.8\%) had experience with 3D-based remote collaboration platforms. Each participant completed all experimental conditions. The study was approved by KAIST's Institutional Review Board (IRB) [KH2024-144].

\subsection{Procedure}
Upon arrival, participants received an overview of the experiment and then completed the informed consent form. 
They then engaged in conversational listening tasks with both male and female avatars under all three conditions (C1–C3).
The order of conditions was randomized using a balanced Latin square design.
After each interaction, participants rated the avatars on Realism, Naturalness, Synchrony of communication cues, and Telepresence. After completing all six sessions (male and female avatars × three conditions), participants identified their most and least preferred avatars and provided reasoning for their choices. Open-ended responses captured detailed feedback on what felt natural or unnatural and what improvements were needed. The full session, including the introduction, lasted approximately 45 minutes.

\subsection{Results}

%\subsubsection{Equivalence Testing}
We used a two-step approach to evaluate differences in participants' ratings for four measures: Realism, Naturalness, Synchrony, and Telepresence. First, we performed a Friedman’s test~\cite{sheldon1996use}, a non-parametric approach, which revealed no statistically significant differences in participant perceptions across the three conditions ($p>.05$ for all measures). 
However, non-significant results from a standard hypothesis test cannot prove that the conditions are equivalent; they only indicate a lack of sufficient evidence for a difference.
To address this, we conducted equivalence testing using the Two One-Sided Tests (TOST) procedure~\cite{lakens2018equivalence} to determine if the differences between conditions were negligibly small.
This is crucial for determining if different data collection methods can produce avatars perceived as equally realistic and natural, even with reduced data and computational demands.
For all equivalence tests, we defined the equivalence bounds ($\Delta$) at $\pm0.5$ on the 7-point Likert scale.
This threshold represents half a scale step, which we considered to be the minimal meaningful difference in user perception~\cite{lakens2017equivalence}.
Any mean difference that falls within this range is treated as negligible, supporting the conclusion that the conditions are practically equivalent. 
The TOST analysis revealed that two conditions, C2 (utterance and emotion data) and C3 (extensive data combination), were statistically equivalent across multiple user perception measures (see Figure~\ref{fig:results}).

For all tests, the conclusion of equivalence was reached because the 90\% confidence interval for the mean difference fell entirely within the predefined equivalence bounds. This result provides strong evidence that reducing data and computational demands (from C3 to C2) does not negatively impact user perception.

\begin{itemize}
    \item \textbf{Realism:} Equivalence was demonstrated between C2 ($M = 3.865$, $SD = 1.093$) and C3 ($M = 3.875$, $SD = 1.202$). The TOST procedure confirmed a negligible difference in perceived realism ($p_{TOST} < .05$).
    The 90\% confidence interval for the mean difference [$-0.567; 0.547$] fell entirely within the equivalence bounds of $\pm0.575$, indicating that users perceive the avatars generated using these two data collection methods as equally realistic.

    \item \textbf{Naturalness:} A similar finding of equivalence was observed between C2 ($M = 3.594$, $SD = 1.158$) and C3 ($M = 3.594$, $SD = 1.039$). The TOST procedure indicated a negligible difference in perceived naturalness ($p_{TOST} < .05$). The 90\% confidence interval for the mean difference [$-0.533; 0.533$] fell within the equivalence bounds of $\pm0.55$, demonstrating that adding extensive data (C3) did not lead to a perceptible improvement in avatar naturalness compared to the more efficient C2.

    \item \textbf{Telepresence:} Equivalence was also confirmed for telepresence when comparing C2 ($M = 4.031$, $SD = 1.155$) and C3 ($M = 4.041$, $SD = 1.174$). The TOST procedure indicated a negligible difference ($p_{TOST} < .05$). The 90\% confidence interval for the mean difference [$-0.575; 0.554$] fell within the equivalence bounds of $\pm0.582$. This implies that users felt equally immersed in telepresence environments when interacting with avatars created using the less data-intensive C2 as they did with those from the more comprehensive C3.
\end{itemize}

\subsection{Qualitative Feedback}
The qualitative feedback provided detailed insights into how participants experienced the avatars’ facial expressions and behaviors, highlighting both engaging features and areas needing refinement.

\subsubsection{Positive Feedback}
Participants consistently reported that the avatars felt more engaging and lifelike when facial expressions, body movements, and speech content were well-aligned. This synchrony helped the avatars appear more human, avoiding the artificial or static quality often associated with avatars.
Subtle facial changes (e.g., cheeks, brows, or forehead) that matched the emotional tone of speech were also appreciated.

\paragraph{Realistic Expression and Movement}
A central source of positive feedback was the avatars’ ability to generate natural and contextually appropriate facial expressions. Participants noted that the avatars’ faces were not static but responded dynamically to conversational cues.
For example, P7 praised the avatar for \textit{``switching between a neutral expression and a gentle smile in response to the conversation,''} noting that this transition felt smooth and credible. Other participants highlighted the avatar's ability to capture subtle details. P5 commented that the avatar \textit{``accurately reflected even micro muscle movements, especially around the eyes and nose,''} which enhanced realism. Similarly, P19 noted that the \textit{``lower-face muscles and the muscles under the eyes moved naturally, making it feel most realistic.''} 
Participants also highlighted nonverbal cues beyond the face, including head and upper-body movements, that mirrored real conversational patterns. P19 observed that \textit{``small forward-back motions that matched the speech made the interaction feel authentic.''} P5 similarly noted that \textit{``the movements of the head and upper body enhanced the realism.''} P12 even observed that the avatar’s body language conveyed mental states, stating that \textit{``the shoulders and head would tilt forward slightly when listening, which helped me feel present in the dialogue.''} 

\paragraph{Synchrony Between Expression and Speech Content}
Another key aspect of positive feedback was the avatars’ ability to align facial expressions with speech content, enhancing emotional congruence.
P8 stated that \textit{``the avatar’s expressions perfectly matched the tone of what was being said, making it feel like I was talking to a real person.''} P14 appreciated that \textit{``even brief utterances triggered subtle but appropriate facial responses,''} highlighting the system’s responsiveness and nuance. P20 mentioned that the avatar \textit{``never felt frozen or out of place during pauses or when switching topics,''} emphasizing the smooth and continuous nature of the expression-speech integration.

Overall, participants emphasized that the combination of accurate micro-expressions, synchronized head and body movements, and context-aware facial responses created an avatar experience that felt immersive, engaging, and convincingly human. These results suggest that careful attention to expression dynamics and behavioral synchrony is critical for enhancing user perception in avatar-mediated communication.

\subsubsection{Areas for Improvement}
Despite these strengths, participants identified three critical areas where avatars fell short: emotional expressiveness, facial coordination, and eye movements.

\paragraph{Emotional Expressiveness and Subtle Movements} Avatars in C1 (Utterance only) were often described as static or expressionless. P6 remarked that \textit{``the avatar’s face appeared unresponsive, which made it feel less realistic,''} and P9 noted \textit{``little variation, making it feel like it wasn’t truly engaging in a conversation.''}
Participants emphasized the importance of subtle emotional dynamics, such as small muscle movements during pauses, to foster empathy and naturalness.
P19 explained, \textit{``subtle movements of the facial muscles during speech pauses would make the avatar feel more realistic, like conversing with a real person.''}

\paragraph{Facial Coordination and Micro-expressions} Several participants stressed the need for smooth coordination across the eyes, cheeks, and mouth to avoid disjointed expressions. Micro-expressions, such as eyebrow twitches or mouth-line shifts, were viewed as especially important for conveying nuanced emotions. P5 highlighted that \textit{``it is important to capture the subtle movements of facial muscles that occur during changes in emotional expression.''}

\paragraph{Eye Movements and Gaze} Eye behavior was repeatedly emphasized as critical for realism and engagement, echoing prior findings on photorealistic avatars~\cite{gafni2021dynamic,zheng2023pointavatar}. P22 noted that \textit{``the sense of communication comes more from whether we made eye contact than from changes in facial muscle movements.''} Others (e.g., P5, P18, P19) reinforced that natural eye contact is \textit{``crucial in creating the feeling that the avatar is actually engaging in a conversation with me.''} Fixed gazes were described as unnatural, with P23 pointing out that \textit{``unconscious movements, such as blinking and subtle shifts, are crucial in creating a sense of realism.''} P16 added that \textit{``when real humans engage in conversation, they naturally focus on the eyes and mouth. The lack of eye movement in the avatar felt unnatural.''}

\paragraph{Uncanny Valley Risks} Participants mentioned that because the avatars were highly photorealistic, even small distortions could be unsettling. P23 described how \textit{``leaning too far forward broke facial proportions, making it look disturbing.''} P15 mentioned glitches that caused \textit{``a cartoon-like distortion,''} and several participants emphasized how minor awkwardness in facial expressions could quickly evoke discomfort. As P1 explained, \textit{``because the avatar closely resembles a real person, it could trigger the uncanny valley effect, making it crucial to ensure facial depiction is as close to real human facial appearance as possible.''} Similarly, P3 stressed that \textit{``Since it’s a photorealistic avatar rather than a standard 3D character, facial expressions become even more crucial. Even the slightest awkwardness can quickly trigger the discomfort associated with the uncanny valley.''}

\section{Discussion}

\subsection{Findings and Implications}

\paragraph{Efficient Data Acquisition for Avatar Generation}

This study demonstrates that photorealistic avatar generation does not necessarily require large-scale data collection. Focusing on conversational contexts, we show that a curated dataset can produce avatars that are both realistic and computationally efficient. In particular, the combination of spontaneous speech (U3) and direct emotional expression (E1) achieved the best balance between quality and efficiency, reducing data requirements by approximately 61\% compared to extensive data collection methods while maintaining comparable performance. These findings challenge the prevailing assumption that exhaustive datasets are essential for high-quality avatars and highlight the value of targeted, context-specific data collection.

\paragraph{Impact of Utterance and Emotional Data Acquisition}
For \textbf{RQ1}, which examines the influence of utterance data acquisition methods on avatar accuracy and efficiency, our results indicate that spontaneous speech (U3) consistently outperformed pronunciation capture (U1) and script reading (U2). While U1 had the lowest computational cost, it produced the least accurate results, whereas U3 offered a well-balanced tradeoff between output quality and computational efficiency, making it the most effective method for capturing utterance data in conversational contexts.

For \textbf{RQ2}, which evaluates the combined role of utterance and emotional data, we assessed six combinations of utterance and emotional data (U1E1, U1E2, U2E1, U2E2, U3E1, U3E2). 
Our analysis revealed that direct emotional expression (E1) combined with U3 outperformed other methods across key metrics (Table \ref{tab:table3}). Based on these insights, we identified U3E1 (spontaneous speech + direct emotional expression) as the optimal combination for subsequent evaluation.

\paragraph{Impact of Data Collection Strategies on User Perception}
Addressing \textbf{RQ3}, we compared three data collection strategies: utterance-only (C1), utterance + emotion (C2), and extensive data combination (C3).
Equivalence testing showed that C2 avatars were rated equally realistic, natural, and immersive to those generated with C3, despite requiring substantially less data and computation. This demonstrates that beyond a certain threshold, additional data yield diminishing perceptual returns, challenging the prevailing assumption~\cite{habermann2020deepcap,zheng2023avatarrex} that extensive datasets are necessary for high-quality photorealistic avatars. 

From a user-experience perspective, these findings emphasize that realism is not linearly correlated with dataset size. Participants’ evaluations revealed that once the core conversational expressions were captured, additional details did little to enhance perceived naturalness or telepresence. Instead, users responded more positively to avatars that conveyed timely, synchronized, and contextually appropriate expressions, which can be achieved with a targeted dataset. 

Taken together, these insights highlight that prioritizing essential expressions enables the generation of avatars that are not only technically feasible but also socially effective. This approach is particularly relevant for AR/VR telepresence, where computational resources and bandwidth are often constrained. By emphasizing efficiency without sacrificing perceptual quality, our framework supports scalable, real-time deployment across platforms and devices, advancing the accessibility of expressive avatars for everyday communication.

\subsection{Limitations and Future Work}
While this study offers valuable insights, several limitations should be addressed in future work. First, the facial expression data and user evaluations were limited to a single cultural group. Because cultural factors strongly influence both the production and interpretation of facial expressions~\cite{masuda2008placing,blais2008culture}, this constrains the generalizability of our findings. Future studies should therefore incorporate data from diverse cultural groups to assess the cross-cultural validity of our methods and improve the robustness of avatar performance.
Second, our user evaluation involved one-way conversational listening tasks with identical stimuli across conditions. Although this design ensured control and consistency, it restricted our ability to examine how avatars facilitate interactive communication. Future research should explore more dynamic, interactive scenarios involving communication to gain a deeper understanding of avatar facial expressions.
In addition, we conducted the study on a large wall display simulating \textit{Project Starline}\footnote{Google Project Starline: \url{https://starline.google/} (Accessed 2026-01-04)}, using pre-recorded Unity scenes instead of real-time rendering. This controlled setup minimized technical issues such as latency and rendering delays, but does not fully reflect real-world XR conditions. Future studies should evaluate performance across different hardware configurations and address constraints such as real-time rendering, bandwidth limitations, and device variability. Such investigations will clarify how effectively photorealistic avatars can be deployed in practical AR/VR applications. Lastly, we plan to prepare an anonymized version of the dataset for public release, enabling broader adoption and facilitating reproducibility.

\section{Conclusion}
This study demonstrates that focusing on essential data inputs, specifically utterance and emotional expression data, can generate realistic avatars with significantly reduced data requirements. 
By targeting expressions most relevant to conversational interaction, we streamlined the data acquisition process, improving efficiency without compromising avatar quality. A key contribution of this work is the identification of core facial expressions for everyday communication, enabling a simplified reconstruction process while preserving high realism and naturalness. We compared three data collection approaches and evaluated their impact on computational performance and user perceptions, including realism, naturalness, and telepresence. Results showed that focused data collection improves performance while maintaining user satisfaction. By integrating quantitative metrics with qualitative feedback, we offer a comprehensive understanding of avatar performance in real-world contexts and provide practical guidelines for efficient avatar creation in AR/VR environments. In conclusion, our streamlined approach to photorealistic avatar generation reduces data demands without sacrificing quality, making high-quality avatars more accessible for real-time communication across devices and platforms. This advances the feasibility of more immersive and engaging AR/VR communication experiences.

\acknowledgments{
This work was supported by the National Research Council of Science \& Technology (NST) (No.CRC21015) and Institute of Information \& communications Technology Planning \& Evaluation (IITP) under the metaverse support program to nurture the best talents (IITP-2025-RS-2022-00156435), both funded by the Korea government(MSIT).
We also acknowledge the support of the Natural Sciences and Engineering Research Council of Canada (NSERC) [RGPIN-2022-03294, ALLRP 590555-23] and Alberta Innovates.}

\bibliographystyle{abbrv-doi}

\bibliography{template}
\end{document}